\documentstyle[12pt,aaspp4]{article}
 
\begin{document}  
 
\begin {flushright} 
OU-TAP-70 \\ 
December 1997 
\end{flushright} 
 
\title{\bf EVOLUTION OF THE POWER SPECTRUM AND THE SELF-SIMILARITY IN
THE EXPANDING ONE-DIMENSIONAL UNIVERSE}  
  
\author{Taihei Yano\altaffilmark{1}
\altaffiltext{1}{Research Fellow of the Japan 
Society for the Promotion of Science.} and Naoteru Gouda}  
  
\affil{Department of Earth and Space Science,  
Graduate School of Science, Osaka University  
Toyonaka, Osaka 560, Japan\\  
E-mail: yano, gouda@vega.ess.sci.osaka-u.ac.jp}  
  
\begin{abstract}  
We have investigated  time evolutions of power spectra of density fluctuations 
for long time after the first appearance of caustics in the expanding 
one-dimensional universe.
It is found that when an initial power spectrum is sale-free 
with a power index $n$,
a self-similarity of the time evolution of the power spectrum is achieved. 
We find that the power spectrum can be separated 
roughly into three regimes according to 
the shape of the power spectrum: 
the linear regime ($k < k_{nl}$ : the regime {\cal 1}),
the single-caustic regime
($k_{nl} < k < k_{snl}$ : the regime 2), and
the multi-caustics regime
($k > k_{snl}$ : the regime 3).
The power index of the power spectrum in each regime has the values of
$n,-1$, and $\mu$ which depends on $n$, respectively.
Even in the case of an initial power-law spectrum with a cutoff scale, 
there might be the possibility of the self-similar evolution of the power 
spectrum after the appearance of the caustics.
It is found, however, the self-similarity is not achieved in this case. 
The shape of the power spectrum on scales smaller than the 
cutoff scale can be separated roughly in two regimes:
the virialized regime ($k_{cut}< k < k_{cs}$ : the regime 4),
and the smallest-single-caustic regime ($ k > k_{cs}$ : the regime 5). 
The power index of the power spectrum is $\nu$ which may 
be determined by the distribution of singular points in the regime 4.
In the regime 5, the value of the power index is $-1$.
Moreover we show the general property about the shape of a power spectrum 
with a general initial condition.
\end{abstract}  

{\it Subject headings}:  
cosmology:theory-large scale structures-selfsimilar
  
\section{INTRODUCTION}  

Formations of the large scale structures in the 
expanding universe is one of the 
most important
and interesting problems of the cosmology.
It is generally believed that these structures have 
been formed owing to gravitational instability.
 Hence it is very important to clarify evolutions of density fluctuations  
by the gravitational instability.
Here we consider density fluctuations of collisionless particles as
dark matters, our interest being mainly
concentrated on effects of the self-gravity.  

The density fluctuations are 
often characterized  by their power spectrum (or two-point 
correlation function).
Thus, it is very interesting and important to investigate time evolutions 
of the power spectrum.  Furthermore, whether the evolutions of 
the power spectrum satisfy the 
self-similarity or not is especially important.
If the self-similarity is satisfied, 
 we can describe completely the time 
evolution of the power spectrum 
without numerical simulations
because 
we can estimate the growth rate of the power spectrum on all scales.
Many authors have investigated the self-similarity of the 
two-point correlation function (or the power spectrum).
Davis \& Peebles (1977) investigated the self-similarity of the two-point 
correlation function by using the BBGKY equations.
It is well known that 
we can make a similarity transformed BBGKY equations
because the gravity is scale-free.
%Thus, it is strongly suggested that there exist a 
%self-similarity of the power spectrum or the two-point correlation function.
They showed the existence of the self-similarity by integrating the 
BBGKY equations numerically 
under the assumptions as shown below.
The  BBGKY equations have the hierarchical structure, that is, 
the time evolution of the $N$-th order correlation function includes
the ($N+1$)-th order correlation function. 
So, the cutoff of the hierarchy is needed
in order to close these equations.
Davis \& Peebles assumed that the three-point correlation function can be 
expressed by the product of the two-point correlation functions and that the
skewness of the velocity field is equal to zero.
Thus the existence of the self-similarity is still uncertain because 
we do not know whether these assumptions are correct or not.
On the other hand, by using $N$-body simulations,
time evolutions of the power spectrum can be calculated directly.
%This is a simple but powerful method.
Recent works, for example, are Colombi,Bouchet \& Hernquist (1996); 
%Yess \& Shandarin (1996); 
Couchman \& Peebles (1997);
 Jain, Mo, \& White (1995) and so on.
They showed that the self-similar evolution of the 
power spectrum can be satisfied when the initial power spectrum is scale-free.
However the relation between the index of the power spectrum and the 
mean relative velocity is still uncertain quantitatively.
Furthermore, only the scale-free cases are investigated.
In this paper we would like to investigate 
not only the scale-free cases but also 
the cases that the initial power spectrum obeys
the power-law with a cutoff.
Hereafter we call this case cutoff-case. 
In the cutoff-case,
the power spectrum obeys the power law even on  
scales smaller than the cutoff scale after the 
first appearance of caustics.
For example, in the one-dimensional system,
The value of the power index is $-1$ 
which can be derived according to the catastrophe theory
(Gouda \& Nakamura (1988,1989)).
Kotok \& Shandarin (1988) also studied the nonlinear spectra 
in the cutoff case and showed that the value of the power 
index is -1.
%Are there  possibilities of the self-similar evolutions after the first 
%appearance of caustics?
Of course, the power spectrum does not evolve in the self-similar form
before the first appearance of caustics.
However it has not been certain whether the time evolution is 
self-similar or not after the first appearance of caustics.
Yano \& Gouda (1997) showed the possibility of the self-similarity in 
this case. 
%and showed that there is the possibility of the self-similar solution 
%which has the index of $-1$ which is decided by the catastrophe theory.
Therefore we investigate the self-similarity not only in the case of 
the scale-free case but also in the cutoff-case.
Then we have to investigate in detail the evolutions of the power spectrum on  
scales smaller than the initial cutoff scale and also those on larger 
scales. Thus we need a wide dynamic range in the wave number space.
However, we cannot get such a wide dynamical range in the 
numerical simulation of three-dimensional 
systems due to the limit of the resolution.
%Especially, when we investigate the self-similarity, 
%the scale grows every moment.
%Hence, we need a wide range of resolution.
Then, as a first step, we consider 
one-dimensional sheet systems.
In the one-dimensional system we can get a wide range 
of resolution for calculating time evolutions of the power spectrum.
%Especially in calculating the case that the initial power spectrum has 
%a power-law with cutoff scale, we need a wide range of resolution.
Furthermore we can have the numerical method for the evolution of
the power spectrum with a good 
accuracy in the 
one-dimensional sheet systems as shown in \S 2.
%Of course, the physical process of the one-dimensional system may be 
%different from that of the three-dimensional system.
%However we believe the strong suggestion about the physical process of 
%the three-dimensional system from the numerical results of the 
%one-dimensional system.
%Furthermore we believe the essence of the physical process 
%is the same the one-dimensional system as the three-dimensional system.
The purpose of this paper is to investigate the time evolution of the 
power spectrum in the one-dimensional systems as a first step. 
We are especially interested in the self-similar evolution of
the power spectrum. 
Therefore, we consider only the background universe which 
does not have a characteristic scale, as a spatial curvature scale.
That is, we consider the Einstein-de Sitter universe.

In $\S 2.1$, we will briefly 
show the numerical method.
We show in $\S 2.2$ and $\S 2.3$
 the numerical results of scale-free case, 
cutoff-case, respectively.
Finally, we will 
devote to the conclusion and discussions in $\S 3$.

%\section{ONE-DIMENSIONAL DENSITY FLUCTUATIONS}
%\subsection{Self-similarity}
%Davis \& Peebles showed the existence of the self-similar 
%solutions under the stability
%condition and the assumptions about the three-point correlation function 
%and the skewness.  In the self-similar solutions, the power index of the 
%two-point correlation function in the strongly nonlinear regime is related 
%to the power index $n$ of the initial power spectrum, 
%where $n$ is defined by  
%\begin{equation}  
% P(k) \propto k^n.   
%\end{equation}  
%The power index of the two-point correlation function in the self-similar 
%solution without assumption of the stability condition
%is derived by using the velocity parameter $h$ 
%(Padmanabhan(1996),Yano \& Gouda (1997)).

\section{NUMERICAL RESULTS}  
%In this section, we calculate roughly two cases of initial conditions.
%One is the scale-free case. This case has been investigated 
%about the self-similarity by many authors.
%Another case is the power-law spectrum with a cutoff scale.
%Even there is a cutoff scale, the power index of the scales smaller than 
%the cutoff scale appears after the first appearance of caustics.
%The value of the index is $-1$.
%So, there is the possibility that the self-similarity 
%exist with the index $-1$ in the non-linear regime.

\subsection{Numerical method}  
We investigate time evolutions of the power spectrum and its 
self-similarity by using a numerical method.
In this paper we consider the one-dimensional system 
that many plane parallel sheets move only a perpendicular direction
to the surface of these sheets.
When two sheets cross, they are allowed to pass through freely each other.
%We investigate mainly after the appearance of the caustics.
%We therefore cannot use the fluid equations. 
%We must solve the Vlasov equation in general.
%However we consider many sheets.
%Each sheet represents a continuous density distribution in the phase space.
 In this sheet system, there is an exact solution
until two sheets cross over as follows
(Sunyaev \& Zel'dovich(1972), Doroshkevich et.al.(1973)):
\begin{eqnarray}  
x &=& q + B_1(t)S_1(q) + B_2(t)S_2(q), \nonumber \\
v &=& \dot{B_1}(t)S_1(q) + \dot{B_2}(t)S_2(q), 
\end{eqnarray}  
where $q$ and $x$ are the Lagrangian and the Eulerian coordinate, respectively.
Here, $S_1(q)$, and $S_2(q)$ are arbitrary functions of $q$. 
$B_1(t)$, and $B_2(t)$ are the growing mode and the decaying mode of linear 
perturbation solutions, respectively.
Since we consider the Einstein-de Sitter universe,
$B_1(t)=a$ and $B_2(t)=a^{-\frac{3}{2}}$,
where $a$ is a scale factor of the universe.
We can compute the crossing time of all neighboring pairs of sheets.
We use a shortest of these crossing time as a time step.
Then, we can compute the new positions and velocities for all sheets 
at this crossing time.
After two sheets cross, we exchange the velocities of just crossed two sheets.
Then we obtain again $S_1(q), S_2(q)$, and therefore exact solutions
as follows:
\begin{eqnarray}  
S_1(q)=\frac{3}{5}a^{-1}(x-q)+\frac{2}{5}\dot{a}^{-1}v, \nonumber \\
S_2(q)=\frac{2}{5}a^{\frac{3}{2}}(x-q)
 -\frac{2}{5}\dot{a}^{-1}a^{\frac{5}{2}}v.
\end{eqnarray}  
These new exact solutions can be used until two sheets cross over.
%We can continue to calculate by repeating the above process.
In this way we obtain the exact loci of the sheets 
by coupling these solutions.
This is the numerical method with good accuracy because 
we connect the exact solutions.
Through this paper, we use $2^{13}$ sheets for numerical calculation.
A period boundary is fixed by a length of $2\pi$. 
Therefore, the smallest wave number is $1$.

\subsection{Scale-free spectrum case} 
In Figs.1(a) and (b), we show the time evolutions of the power 
spectrum with the scale-free 
initial power spectrum given by
\begin{equation}  
P(k,t_{ini}) \propto k^n,
\end{equation}   
where 
$n$ is the power index of the initial spectrum.
It is found that the following results are satisfied for 
$-1 < n \leq 4$ .
The condition of $n > -1$ means the hierarchical clustering in the
one-dimensional system.
We are interested in this case ( $n > -1$ ) for ``scale-free'' case.
One of the reasons why we consider this case is that the hierarchical 
clustering picture is expected in the real world. 
Another reason is that the results for $n < -1$ is similar to those in 
the case of the single wave case.
See also eq.(14).
%( i.e. $P(k=1,t_{ini}) \neq 0$, $P(k \neq 1,t_{ini})=0$).
Then the results for $n<-1$ can be reffered from the results 
shown in the regime4 and 5 in the ``cutoff case''(see $\S 2.3$).
The condition of $n \leq 4$ is required because we consider
the situation in which the non-linear mode coupling from 
higher $k$ to smaller $k$ can be neglected and so the linear perturbation 
theory can be hold on small $k$( Peebles (1980), 
Shandarin \& Melott (1990), Gouda (1995)).
In the following, we show the case of $n=1$ and $2$ for example.
 Fig.1(a) and(b) are the $n=1$ case and the $n=2$ case, respectively.
We are interested in the hierarchical clustering picture.
Therefore we chose some examples that have the power index larger than $-1$.
We can obtain the same results qualitatively in the cases of the other indexes.
Here $t_{ini}$ is the initial time, and the initial scale factor of 
these two cases are $0.1$ and $0.01$, respectively.
We normalize the scale factor as follows; $a=1$ when the first caustic 
has appeared.
The phases of the initial fourier spectrum are given in random.
%like a random gaussian. However, an amplitude is fixed in proportion
%to the expected value.
We have averaged 50 samples.
%We have taken 50 times sample average. 
Here, we scale the wave number and power spectrum as follows: 
%in order to ascertain whether each power spectrum coincides.
%Therefore, we define the following scaling wave number $k_*$
%and scaling power spectrum $P_*$
\begin{equation}  
k_* \equiv \frac{k}{k_{nl}(t)},
~~~~P_*(k_*,t) \equiv \frac{P(k,t)}{P_{scale}(t)}, 
\end{equation}  
where $k_{nl}$ is defined by
\begin{equation}  
\frac{1}{2\pi}\int^{k_{nl}(t)}_{0}P(k,t)dk =1.
\end{equation}  
In the regime of $k<k_{nl}$, the power spectrum grows according 
to the solution of linear perturbation, that is,
the power spectrum satisfies the relation $P(k,t) \propto a^2$.
Therefore, $k_{nl}$ is proportional to $a^{-2/(n+1)}$.
$P_{scale}(t)$ is defined by 
\begin{equation}  
P_{scale}(t) \equiv P(k_{nl}(t),t).
\end{equation}  
Then $P_{scale}(t)$ is proportional to $a^2 k_{nl}^n \propto a^{2/(n+1)}$.
This scaled power spectrum $P_*(k_*)$ 
is shown in Fig.2(a) for the case of $n=1$ . 
The same one but for the case of $n=2$ is shown 
in Fig.2(b).
We can see the coincidence of each power spectrum
at each time with good accuracy.
This means that the self-similarity is achieved in the scale-free case. 
Furthermore we can see the three different power indexes in three regimes:
the linear regime ($k<k_{nl}$ : the regime 1), the single-caustic regime
 ($k_{nl}< k <k_{snl}$ : the regime 2), and 
the multi-caustics regime 
 ($ k >k_{snl}$ : the regime 3).
Fig.3 shows 
the schematic general power spectrum at a time in the scale-free case.
The value of the power index in the linear regime, of course,
remains unchanged.
We explain the power index of the power spectrum in the other 
two regimes: the single-caustic regime, and the multi-caustics regime 
in the following subsections.

\subsubsection{Single-caustic regime (the regime 2)} 
In the single-caustic regime 
($k_{nl}< k <k_{snl}$ : the regime 2)
we can see that the power index value becomes $-1$.
The reason is as follows;
The density perturbation with the scale which just entered the 
non-linear regime at a given time, that is, $k_{nl}$, makes ``caustics''.
Strictly speaking, the caustics would be appeared if the initial spectrum was 
smoothed bellow the scales just entered the non-linear regime.
Therefore, we must notice that the real caustics cannot be observed
in the scale-free case 
because of smearing by the small scale fluctuations. However we call 
these scale on $k_{nl} < k < k_{snl}$,
 "single-caustics regime" in this paper. 
After the first appearance of "caustics", 
the value of the power index of the power 
spectrum on the scales smaller than that scale is $-1$
(Gouda \& Nakamura (1989), Kotok \& Shandarin (1988)):
Here we briefly show why the value of the power index is $-1$ 
after the first appearance of caustics.

The fourier spectrum $\delta_k$ of the density fluctuation is given by 
\begin{equation}  
\delta_k =\int \delta(x) e^{ikx}dx.
\end{equation}  
Then the density is given by 
\begin{equation}  
\rho (x)=\frac{\rho_0}{|\frac{dx}{dq}|},
\end{equation}  
where $\rho_0$ is the mean density of the universe.
At the Lagrangian singular point $q_0$, the following relation is satisfied.
\begin{equation}  
(\frac{dx}{dq})_{q_0}=0.
\end{equation}  
At the singular point $q_0$, the density $\rho (x)$ diverges.
Therefore, the Eulerian coordinate $x$ can be written 
\begin{equation}  
x=x_0 + \beta (q-q_0)^2 +O((q-q_0)^3).
\label{7}
\end{equation}  
For simplicity, we can put $x_0 =q_0 =0$ without losing generality.
Therefore we can express $x=\beta q^2$ around the caustic. 
The density around the caustic is expressed by
\begin{equation}  
\rho (x)=\rho_0|\frac{dx}{dq}|^{-1}\propto (\beta x)^{-\frac{1}{2}}
\label{8}
\end{equation}  
This density profile determine the proper index of the power spectrum.
This type of singularity is called A2 type according to the 
catastrophe theory (Gouda \& Nakamura (1988,1989)).
Eqs.(\ref{7}), and (\ref{8}) can be generally satisfied around the singular 
points in the multi-stream flow regimes where the Zel'dovich 
solution cannot hold.
The A2 type of singularity is the only stable type in the 
one-dimensional system and their eqs.(\ref{7}), and (\ref{8}) remain 
in the multi-stream flow regions after the first appearance of caustics.
Here it must be noted that Lagrange coordinate $q$ in the above 
argument is not the initial position of the sheet.
Please refer to Roytvarf (1987), and Gouda \& Nakamura (1989), 
for the detailed explanation.
On the other hand, 
it is found that $\rho(x)$ is proportional to $x^{-2/3}$ 
around the singular points at the first appearance of caustics
(Zel'dovich (1970), Arnold et.al.(1982), Gouda \& Nakamura (1988)).
This type of singularity is called A3 type. 
The A3 Type is not structural stable in the one-dimensional system 
and then the A3 type of singularities appears only at an instant.
It disappears and quickly evolve into the A2 type, i.e. 
$\rho \propto x^{-1/2}$ singularity.
Hence the contribution from the A3 type singularity in estimating the 
power spectrum is negligible.
The fourier spectrum of the density is given by
%Even in the multi-stream flow regime, 
%the density profile remain $\rho \propto x^{-1/2}$ as in the 
%Zel'dovich approximation when we consider the regime smaller than the 
%caustic separation (Roytvarf(1987)).
%In more realistic approach the singularities do not 
%occur and the maximum density is not so high in, for example, the neutrino 
%dominated model because of its velocity dispersion
%(Zel'dovich \& Shandarin (1982),Kotok \& Shandarin (1987)).
\begin{eqnarray}  
%\delta_k &=& \int \delta(x) e^{ikx}dx \nonumber \\
%         &=& \int e^{ik\beta q^2}dq \nonumber  \\
%         &=& (k\beta)^{-\frac{1}{2}}\int e^{it^2}dt,
\delta_k &=& \int \delta(x) e^{ikx}dx \nonumber \\
         &=& \int \frac{\rho}{\rho_0}e^{ikx}dx \nonumber  \\
         &\propto& \int (\beta x)^{-\frac{1}{2}} e^{ikx}dx\nonumber  \\
         &=& \beta ^{-\frac{1}{2}}k^{\frac{1}{2}}k^{-1}
\int t^{-\frac{1}{2}}e^{it}dt,
\end{eqnarray}  
where 
%we used the relation of mass conservation, i.e. $\rho(x)dx=\rho_0 dq$.
%Here 
%$t^2 \equiv k \beta q^2$.
$t \equiv kx$.
Then, we obtain
\begin{equation}  
P(k) \propto (k\beta)^{-1}.
\label{10}
\end{equation}  
From eq.(\ref{10}), we find the value of 
the power index is $-1$.
%You may think that the A3 type singularity, that is, $\rho \propto x^{-2/3}$
%singularity appears at first time.
%However, this type of singularity quickly evolve into $\rho \propto x^{-1/2}$
%singularity.  Indeed, the measure of the A3 type singularity is zero.  
%Therefore 
%we do not have to consider the A3 type singularity.
Here we notice that small $\beta$ contributes to the large amplitude of the
power spectrum.
Furthermore, the coefficient $\beta$ is very small when the caustics have 
just appeared.
We can see this fact 
by the evolution of the 
``single-wave perturbation''.
Here, the single-wave perturbation means the density fluctuation whose 
initial condition is given by
\begin{equation}  
x(q)=q + X sin(q),
\end{equation}  
where $X$ is a constant value.
%The density fluctuation are given by a cosine wave whose wave length
%is a size of a period boundary.
%Peculiar velocities are given so as to have the only growing mode, that is 
%so called the Zel'dovich approximation.
Hereafter we call this case the single-wave case.
The first caustics have just appeared at a center 
of the $x$ axis ($x=0$) in Fig.4(a). 
After the first appearance of this caustic, more caustics appear
in the collapse regime by phase mixing (Fig.4(b)-(d))
(Doroshkevich et.al(1980), Melott (1983), Gouda \& Nakamura (1989)).
A caustic (singularity of the density field) is located at the point 
where the derivative of $v$ with respect to $x$ is infinity.
The absolute value of the derivative of $v$ with respect $x$ around the 
singular points is proportional to $\beta ^{-1}$. Then as the phase mixing 
continues and so more caustics appear, it is found from Fig.4(a)-4(d)
that $\beta$ increase because $|\frac{dv}{dx}|$ around the singular 
points decrease.
Hereafter we call the structure in the phase space shown in Fig.4 (a)-(b)
``the whirlpool''.
Just after the multi-caustics have appeared, 
the coefficient factor $\beta$ of this wavelength 
is very small and the effect of the 
amplitude of the power spectrum in these waves dominates.
As a result, the index of $-1$ that is predicted by the catastrophe theory
appears. Therefore we call these regimes, single-caustic regime.

Here, we comment that in the $\it real$ singularities do not occur in 
the real world.
Of course, the arguments on the caustics in this paper cannot be applied 
to barionic gas because shocks appeared and they prohibited that the 
density diverges.
Furthermore the maximum density is surprisingly low even in the 
neutrino-dominated model due to their velocity dispersion
(Zel'dovich \& Shandarin (1982),Kotok \& Shandarin (1987)).
In this paper, we consider the cold collisionless matters 
whose velocity dispersion is much small compared with the 
scales under consideration.

\subsubsection{Multi-caustics regime (the regime 3)} 
In the multi-caustics regime ($k>k_{snl}$), we can see the power law spectrum.
With the power index of these regimes is different from $-1$ predicted 
by catastrophe theory.
In this regime, various small scale fluctuations ($k \gg k_{snl}$)
have already collapsed and made singularities.
Every singularity makes ``the whirlpool'' in phase space as shown 
in the single-wave case (Fig.4).
Various size of ``whirlpools'' are made of the various 
scale of fluctuations.   
The distribution of the whirlpools determine the value of the power index.
This distribution depends on the initial power index $n$.
That is, the power index $\mu$ in this regime depends on $n$.
Indeed, the value $\mu$ depends on the initial 
power spectrum as shown in Fig.2(a), and 2(b).
Therefore we call this regime the multi-caustics regime.
When the self-similarity is satisfied, there is a relation between 
the power index of the 
two-point correlation function, the initial power spectrum index, and the 
mean relative peculiar velocity.
Here, we briefly show this relation 
in the $w$ dimensional system.
In the linear regime of the $w$ dimensional system, the two-point 
correlation function is given by(Peebles 1980,1993),  
\begin{equation}  
\xi \propto a^2 x^{-(w+n)} \propto (\frac{x}{a^\alpha})^{-(w+n)},  
\end{equation}  
where  
\begin{equation}  
\alpha \equiv \frac{2}{w+n}.  
\end{equation}  
On the other hand,   
in the strongly nonlinear regime, the two-point correlation function 
obeys the following evolution equation 
(0th moment equation of the 2nd BBGKY eq.:  
Davis \& Peebles (1977), Yano \& Gouda (1997)),
\begin{equation}  
   \frac{\partial \xi}{\partial t}  
+ \frac{1}{ax^{(w-1)}}  
\frac{\partial}{\partial x}[x^{(w-1)} (\xi+1)\langle v \rangle]=0.  
\label{bbgky}
\end{equation}  
In the strongly non-linear limit, we can assume a power law solution 
for the two-point correlation function (Yano \& Gouda (1997)).

\begin{equation}  
\xi \propto a^{\epsilon}x^{-\gamma}.
\end{equation}  
In this case, we obtain 

\begin{equation}  
\langle v \rangle=-h\dot{a}x,~~~~~(\epsilon-h(w-\gamma)=0)
\end{equation}  
and then,
\begin{equation}  
\xi \propto a^{h(w-\gamma)}x^{-\gamma}= (\frac{x}{a^{\alpha '}})^{-\gamma},  
\end{equation}  
where $h$ is the constant and
\begin{equation}  
\alpha ' \equiv \frac{h(w-\gamma)}{\gamma}.  
\end{equation}  
Hereafter we call the parameter $h$ the relative velocity parameter.
If the self-similarity is satisfied, $\alpha '$ must be equal to $\alpha$.  
Then the power index $\gamma$ is given by  
\begin{equation}  
\gamma=\frac{wh(n+w)}{2+h(n+w)}.  
\label{11}
\end{equation}  

  If the relative velocity parameter $h$ is $1$ (stability condition)
as Davis \& Peebles (1977) assumed,
we obtain the $\gamma=(n+1)/(3+n)$ in the one-dimension ($w=1$). 
In this case, the power index 
%$\gamma_k \equiv \gamma -w$ 
$\mu$ of the power spectrum is given by 
$\mu=\gamma -w=-2/(3+n)$.
However the power index 
$\mu$ which is given by the  numerical 
results of the power spectrum 
is different 
from $-2/(3+n)$.
Therefore, the stability condition ($h=1$) is not satisfied.
%Caustics with various waves appear as time goes on.
%When we consider a regime, informations of the density fluctuation 
%in this scale are important in general. In these regime, distribution 
%of caustics is more important than the shape of $\it one$ caustic.
%Because the characteristic separation of caustics is smaller than the scale 
%of this regime. These caustics appear from the various scale waves and the 
%relation between each amplitude of waves depends on the initial condition. 
%Then, these distributions depend on the initial conditions. Therefore the 
%index of these scales has a certain initial dependent value $\mu$.
We can estimate the velocity parameter $h$ from the power index 
by using eq.(\ref{11}).

\begin{equation}  
h=\frac{2\gamma}{(n+w)(w-\gamma)}=-\frac{2(\mu+w)}{(n+w)\mu}
=-\frac{2(\mu+1)}{(n+1)\mu}.
\end{equation}  
%where
%\begin{equation}  
%\gamma_k=\gamma-w~~~~~~~~ (P(k) \equiv k^{\gamma_k} : k \gg k_{snl}). 
%\end{equation}  
We show the results in the Table 1.
Yano \& Gouda (1997) discussed that the velocity parameter $h$ has the 
value between $0$ and $1$. Indeed the value of $h$ stays at this range.
We show the numerical result of $h$ not only for the $n=1$, and $2$ cases 
but also for the $n=0$, and $3$ cases.

%The value of $h$ is different from $1$. The stable clustering is not 
%satisfied. In the small $n$ cases, when the smaller objects have clustered,
%small clustered objects can easily merged and are easy to virialize because 
%the larger merged object appears a little time after the smaller objects have 
%clustered. So, the parameter $h$ becomes smaller. In the large $n$ cases, 
%when the smaller objects have clustered, small clustered objects are more 
%difficult to merge or virialize because the larger merged object appears a 
%longer time after the smaller objects have clustered than in the small $n$ 
%case.  So, the parameter $h$ becomes larger than with the small $n$ case.
%On the other hand, Yano \& Gouda discussed the parameter $h$ and showed that 
%the parameter have the value of order $1$.  In the  hierarchical clustering, 
%there are two extreme situations in the way of clustering.  In this first 
%one, the collapsed object cannot be broken and clustered together to form 
%the larger cluster.  In this situation, the mean separation of the particles 
%does not change as time increases, and $h=1$. This corresponds to the 
%stability condition (stable clustering).  In the other one, the smaller 
%objects have clustered and merged together and the completely virialized 
%object is newly formed. In this case, the mean separation of the particles is 
%expanding with the Hubble velocity and then $h=0$ (comoving clustering). So, 
%the parameter $h$ becomes the indicator of the merging rate of the clusters. 
%Therefore, these results are consistent with the discussion of Yano \& Gouda.

Table 1. $n$ dependence of $\mu$ and $h$

\tabcolsep=10mm
\begin{tabular}{ccc}
$n$ & $\mu$ & $h$ \\ \hline
0 &-0.88 &0.29\\
1 &-0.75 &0.33\\
2 &-0.65 &0.36\\
3 &-0.58 &0.37
\end{tabular}

\subsection{Cutoff-case}
We consider the following initial spectrum, that is, the 
cutoff-case.
%\begin{eqnarray}  
%P(k)&\propto& k^n (k<k_{cut}~:~ n=1,2)\nonumber \\
%P(k)&=&0 (k>k_{cut})
%\end{eqnarray}  
\begin{equation}  
P(k)\left\{\begin{array}{ll}
\propto k  &  (k<k_{cut})\nonumber \\ 
=0 & (k>k_{cut})
\end{array}\right..
\end{equation}  
The time evolution of the power spectrum for the cutoff-case is shown in Fig.5.
It is shown only for the $n=1$ case. However, we obtain essentially the 
same result for other cases ($n=2,3$, and $0$).
As we can see from Fig.5, until the first appearance of caustics (lowest solid 
line of the power spectrum except for initial spectrum in Fig.5), 
the amplitude of the power spectrum grows even on scales smaller than the 
cutoff scale.
The power index of these regime is predicted to be $-1$.
Because the wave number $k=k_{cut}$ makes the caustic and this caustic 
results in the predicted value of power index as shown in $\S 2.2$.
 Indeed, we find that the value of the index 
in these regimes is nearly equal to $-1$.
The scaled power spectrum is also shown in Fig.6.
The definition of the scaling is the same as that in the scale-free case. 
As we see from Fig.6, each power spectrum at the different time
 does not coincide.
Therefore the self-similarity is not satisfied.

%This large growth rate is understood in the following way.
%When the first caustic appears, the power index 
%in the scales smaller than the 
%cutoff scale appears. After that, more and more caustics of larger scale
%(smaller wave length) appear one after another. Therefore, each caustic 
%contributes to the amplitude of the power index.

The characteristic scale of  separation of the caustics becomes smaller 
and smaller 
as time increases (refer to the case of the single-wave: See Fig.4 (a)-(d)). 
Here, $k_{cs}$ is defined as the wave number of the characteristic scale.
On the scales smaller than the characteristic scale 
$k>k_{cs}$(regime 5: See Fig.10), the power index of the 
power spectrum becomes $-1$.  
Because on the scales smaller than the characteristic scale, the 
density profile around one singular point determines the power spectrum.
Therefore we call this regime the smallest single-caustic regime.
On the other hand, the regime $k_{cut}<k<k_{cs}$ (regime 4), 
the power index of the 
power spectrum is different from $-1$.  
The index has a certain value $\nu$.
Because on the scales larger than the characteristic scale, 
smoothed density profile with the smoothing scale ($k_{cut}<k<k_{cs}$)
determine the power spectrum.
We call this regime the virialized regime.
We show the schematic of the density distribution in Fig.7.
We notice that the wave number $k_{cs}$ which represents the
characteristic separation of caustics becomes larger and larger as time 
increases.
On the contrary, the non-linear scale wave number $k_{nl}$ and $k_{snl}$
become smaller and smaller as time increases.
Thus,
the power spectra of the different times cannot coincide on all scales.
Therefore the self-similar evolution in the all regimes including
the scales smaller than the 
cutoff scale cannot be satisfied.

We compare the time evolutions of the 
power spectrum for the  cutoff-case with those for the scale-free case 
in Fig.8.
Of course, the power spectrum for these two cases does not coincide 
with each other on all scales.
On the other hand, on the scales larger than the cutoff scale, 
we can see the coincidence of  power spectra for these two cases.
This means that the growth rate on the wave number smaller than $k_{cut}$
does not affect the difference of the larger wave number than the cutoff scale.
%Therefore we can consider separately the scale smaller than the cutoff scale
%and the scale larger than the cutoff scale.
%On the other hand, on the scales larger than the cutoff scale, 
%time evolution of the power spectrum is almost the same as that in the 
%scale-free power spectrum case, in spite of the existence of mode coupling 
%with all scales by the non-linearity of a growth of fluctuations.
%That is, on the scales larger than the cutoff scale, the effect of 
%the mode coupling with smaller scales than the cutoff scale is negligible.  
%If we neglect scales smaller than the cutoff scale, 
%the evolution of the power spectrum in the initial power-law 
%spectrum with the cutoff scale coincide with the scale-free case.
%This means that the self-similar evolution is almost satisfied 
%in the scale larger than the cutoff scale.
We show the scaled power spectrum for the 
cutoff-case only on the wave numbers smaller than the cutoff scale 
$k_{cut}$ in Fig.9.
This means that even in the cutoff-case, the self-similar evolution 
can be satisfied on scales larger than the cutoff scale.

\section{CONCLUSIONS AND DISCUSSION}  
We have calculated the time evolution of the power spectrum for
two cases of initial conditions.
One is the scale-free case and the other one is the cutoff-case.
In the case of the scale-free case, we can see the self-similar 
evolution of the power spectrum. The scaled power spectrum 
$P_*(k_*)$ at each time coincides with each other.
We can separate roughly the power spectrum into three regimes.
One is the linear regime ($k < k_{nl}$ : the regime 1).
The value of the power index in this regime 
remains the initial power index, $n$.
The second regime is 
the single-caustic regime ($k_{nl} < k < k_{snl}$ : the regime 2).
In this regime, the power index becomes $-1$ and is independent of the 
initial conditions. This result is caused by the 
appearance of caustics at this scale, and these caustics determine 
the power index of the power spectrum in this regime.
The third regime is the multi-caustics regime ($k > k_{snl}$ : the regime 3).
%This regime has the smallest scale regime in the scale-free case.
The distribution of the ``whirlpool'' in phase space 
determine the value of the 
power index.
Therefore in this scale the power index $\mu$ has the value which depends 
on the initial condition.
%This value is different from the value of the power index in the 
%single-caustic regime ($k_{nl} < k < k_{snl}$ : the regime 2).
We can estimate the velocity parameter $h$ from the power index in the 
multi-caustics regime, which is around 0.5.
Yano \& Gouda (1997) discussed the probable value of $h$ from the physical 
point of view, and obtained that $h$ takes a value between $0$ and $1$.
 Estimated values of $h$ are consistent with this argument.
Indeed, the index $\mu$ and the velocity 
parameter $h$ depend on the initial power index $n$.
The stability condition ($h=1$) is not satisfied in this case.

In the cutoff-case,
we find that there is no self-similarity on all scales.
The scaled power spectrum does not coincide with each other in 
all regimes. 
However, the spectrum coincides on the scales larger than the cutoff scale.
After the appearance of the first caustics, the power index in the regime 
of the scale smaller than the cutoff scale becomes $-1$.
This value is, as mentioned above, caused by the appearance 
of caustics as shown in Fig.3.
More and more caustics appear one after another and so the separation of 
caustics becomes smaller and smaller.
Even after the appearance of many caustics, on scales smaller than the 
characteristic separation of the caustics ($k > k_{cs}$ : the regime 5), 
the power index 
of the power spectrum is obtained by the 
density profile around the singular point.
Therefore the power index of these scales becomes $-1$
which can be derived according to the catastrophe theory.
On the other hand, on the scales larger than the characteristic
separation of the caustics ($k_{cut} < k < k_{cs}$ : the regime 4),
 the distribution of the singular points
determines the power index on these scales instead of the density 
profile around one singularity.
%The shape of one caustic is less important.
Because in this regime, the smoothed density profile with the smoothing 
scale ($k_{cut}<k<k_{cs}$) determine the power index. 
On these scales, the distribution of the singularity occurring in the 
evolution of the single-wave is important.
Then, the power index on these scales is determined by this distribution 
of the singularity.
%The power index on these scales is determined by the value which 
%can be derived by the distribution of the singularity occurring in the 
%evolution of the single-wave.
Therefore, we can roughly separate two regimes on the scales smaller than the 
cutoff scale $k_{cut}$.
One is the virialized regime ($k_{cut} < k < k_{cs}$ : the regime 4),
and the power index has the value $\nu$ .
Another is the smallest single-caustic regime ($k > k_{cs}$ : the regime 5) 
and the index have the value of $-1$.
Both indexes are independent of the initial conditions.

In a real situation about the evolution of the power spectrum, 
the initial power spectrum has a cutoff at a certain scale. 
Therefore, in the case of a real situation, 
we can consider the evolution of the cutoff-case. 
We notice that even if the initial power spectrum does not have a cutoff,
it decreases with the power index $n<-1$ on certain scales ($k>k_{dec}$).
Therefore the time evolution of the power 
spectrum is the same with the one in the cutoff-case
after the appearance of caustics.
And then, we can separate roughly five regimes.
We show a schematic of the power spectrum at a time after the 
first appearance of caustics in Fig.10.
First regime is the linear regime ($k < k_{nl}$ : the regime 1). 
The index of the power spectrum in this regime is $n$.
%Of course this index depends on the initial conditions. 
Second regime is the single-caustic regime 
($k_{nl} < k_{snl}$ : the regime 2). 
The index in this regime is $-1$, and independent of the initial conditions.
Third regime is the multi-caustics regime 
($k_{snl} < k_{cut}$ : the regime 3). 
The power index is $\mu$ which depends on the initial power index $n$.
Fourth regime is virialized regime 
($k_{cut} < k_{cs}$ : the regime 4). 
The value of the power index is $\nu$ which is independent of the initial 
power index.
The fifth regime is the smallest single-caustic regime 
($k > k_{cs}$ : the regime 5). 
The value of the power index is $-1$.
Only when we consider the evolution of the power spectrum on 
scales larger than the cutoff scale (i.e. the first, 
second, and third regime), the self-similarity is satisfied.

%When we consider the self-similarity, we used the power spectrum instead of 
%the two-point correlation function. These two functions are equivalent
%about the information of the density fluctuation because there are 
%related by the following form:
%\begin{equation}  
%\xi=\frac{1}{2\pi}\int P(k)e^{ikx}dk
%\label{xi}
%\end{equation}  
%However, we have a finite dynamic range.
%We do not have the power spectrum of small scales.
%Even there is no power spectrum in small scales, we can understand that 
%the power spectrum is almost correct by the same discussion in Fig.8.
%On the other hand, the amplitude of $\xi$ is not described correctly in the 
%non-linear regime because the integral of the power spectrum is finite 
%regime (eq.(\ref{xi})). 
%So, it is not recommended to investigate the self-similarity by using the 
%two-point correlation function.
%Furthermore, we estimated the parameter $h$ from the power index of the 
%power spectrum instead of estimating $h$ directly from the mean relative 
%velocity, the reason being that as we can see from eq.(\ref{bbgky}),
%$\langle v \rangle$ is not good in the non-linear regime. 

\acknowledgments  
  
We would like to thank T.Tanaka for important comments.
And we would like to thank 
E. Van Drom for useful suggestions.
We are grateful to the referee, S.F. Shandarin for many
useful and important suggestions.
 This work was supported in part by 
Research Fellowships of the Japan Society for the Promotion of Science 
for Young Scientists (No.4746).

\newpage

\noindent
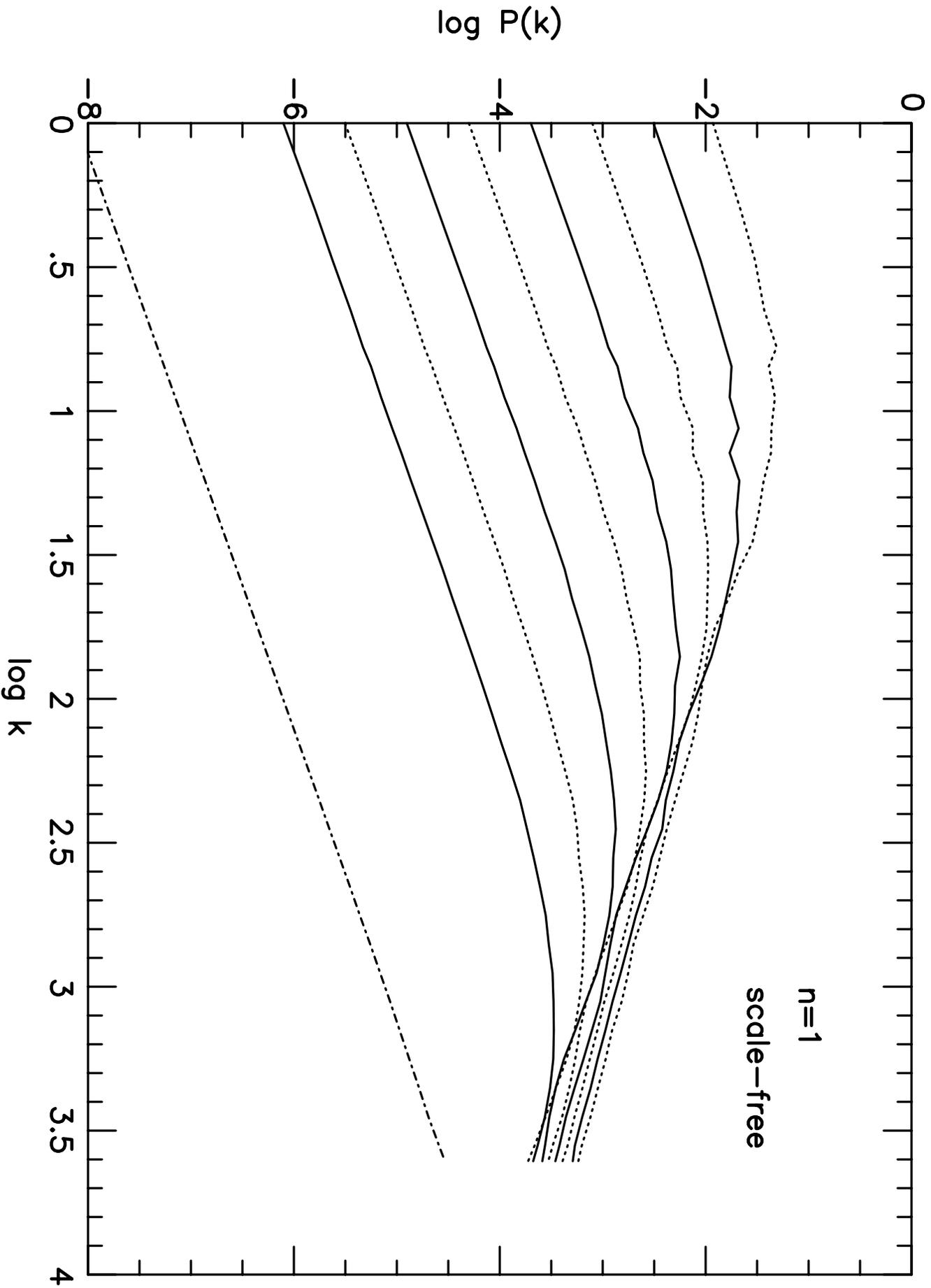
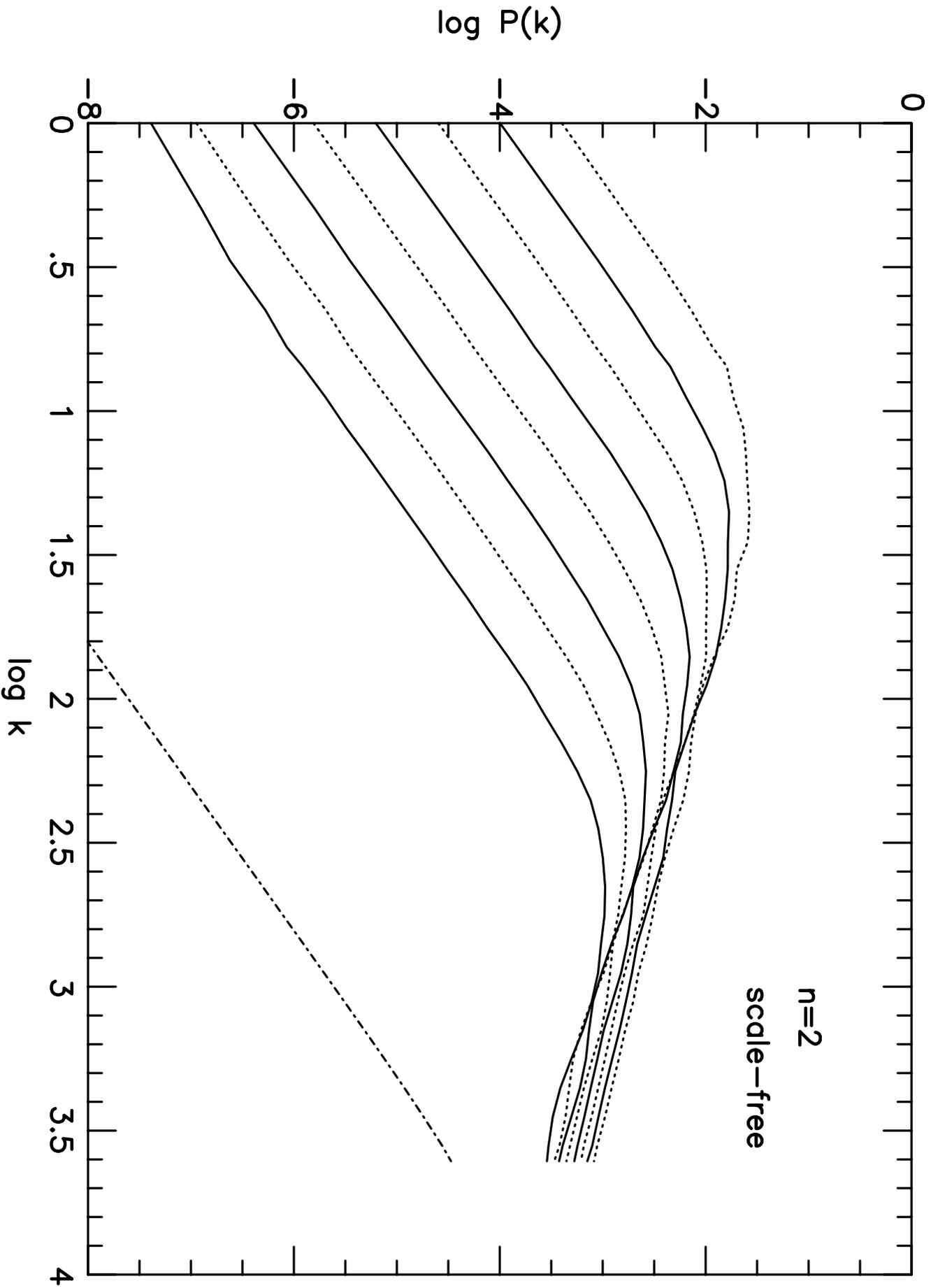
Fig.1.-(a) The time evolution of the power spectrum. 
The 
solid and dotted lines are used mutually in order to distinguish easily 
each line.
The curves shown are (bottom to top)  at $a=0.1,1,2,4,8,16,32,64,128$.
The initial power spectrum is scale-free with the power 
index $n=1$(Dotted-dashed line).

\noindent
(b) The same as (a) but for the initial power index $n=2$ and the power 
spectra are shown at the scale factor $a=0.01,1,2,4,8,16,32,64,128$.

\noindent
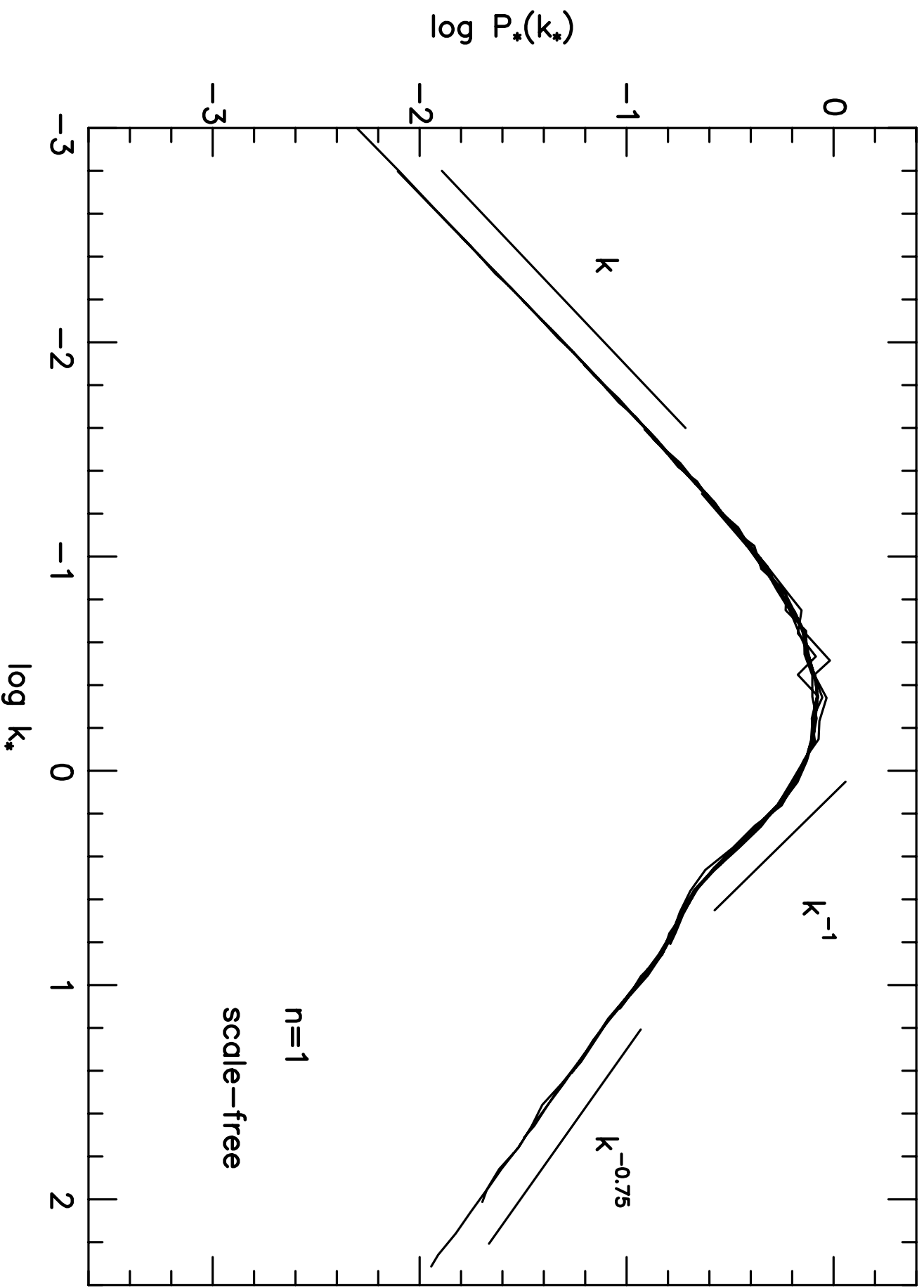
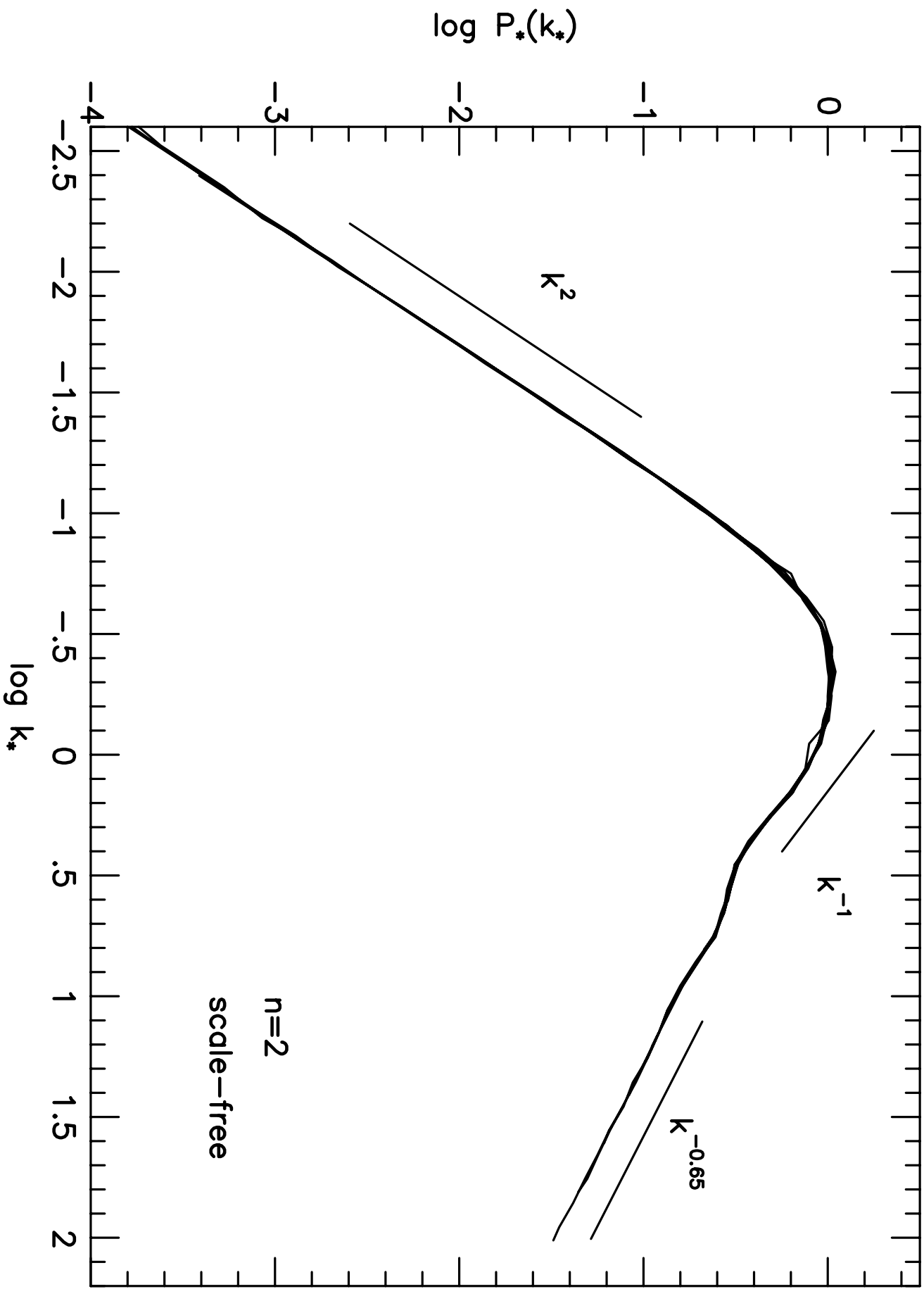
Fig.2.-(a) The scaled power spectrum $P_*(k_*)$.
The initial power spectrum is scale-free with the power index $n=1$.
Three solid straight line shows the power law with the power index 
of  1,-1,-0.75.

\noindent
(b) The same as (a) but for the initial power index $n=2$.
Three solid straight line shows the power law with the power index 
of  2,-1,-0.65.

\noindent
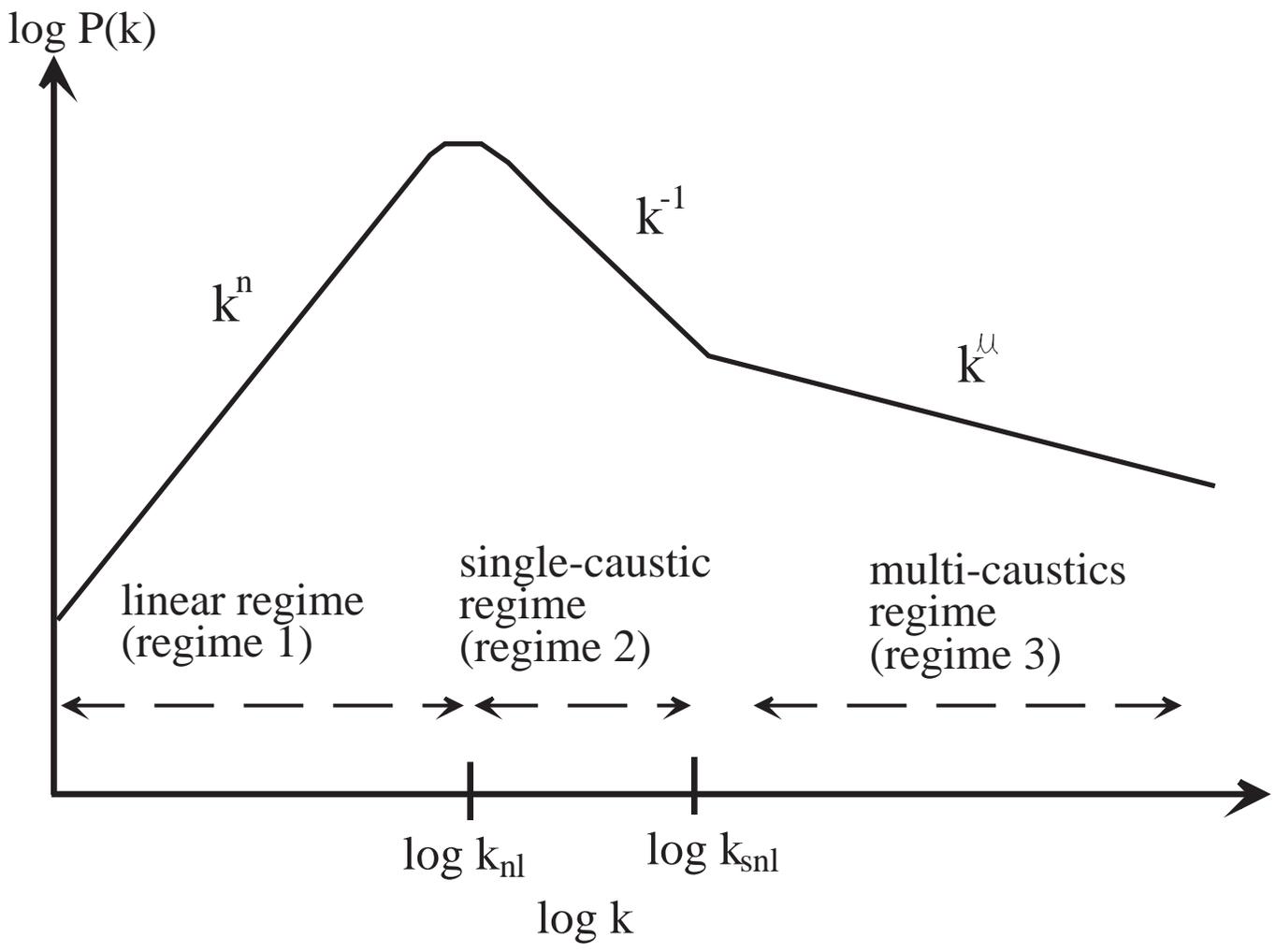
Fig.3.- The schematic of the power spectrum in the scale-free 
case at a certain time.

\noindent
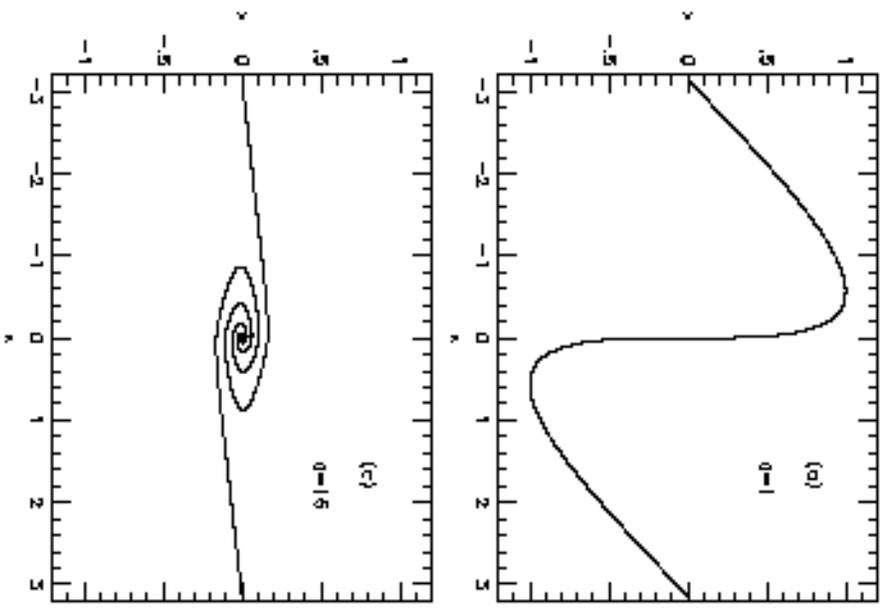
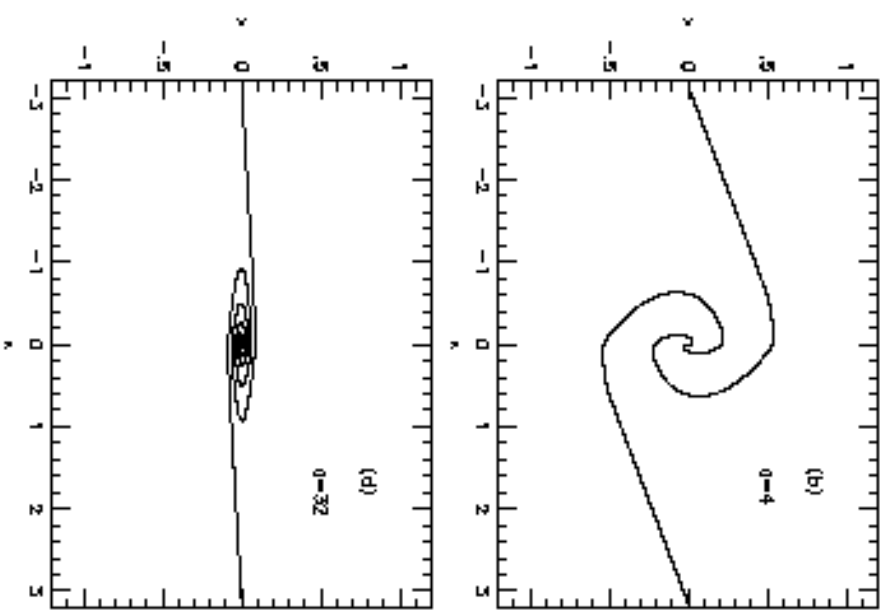
Fig.4.-(a) The distribution in the phase space for the single-wave case at the 
first appearance of caustics. A horizontal axis and a vertical axis are 
the Eulerian coordinate $x$ and the velocity, respectively. 
Scale factor $a$ is normalized at this time ($a=1$).

\noindent
(b) The same as (a) but at $a=4$.

\noindent
(c) The same as (a) but at $a=16$.

\noindent
(d) The same as (a) but at $a=32$.

\noindent
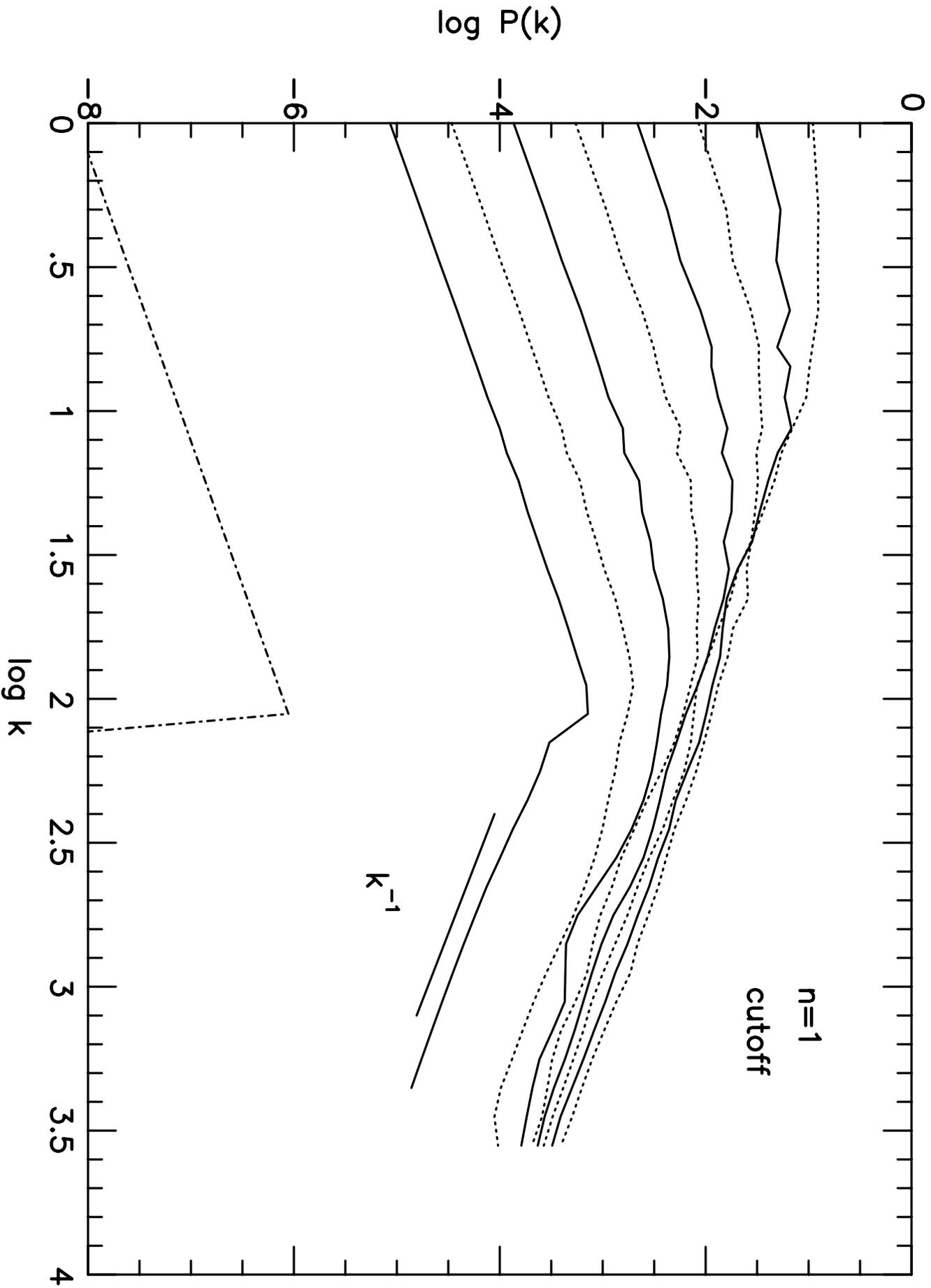
Fig.5.- The time evolution of the power spectrum.
The solid and dotted lines are used mutually in order to distinguish easily 
each line.
The curves shown are (bottom to top) at $a=0.05,1,2,4,8,16,32,64,128$.
The initial power spectrum obeys the power-law with the index $n=1$ 
and it has a cutoff at $k_{cut}=127$(Dotted-dashed line). 

\noindent
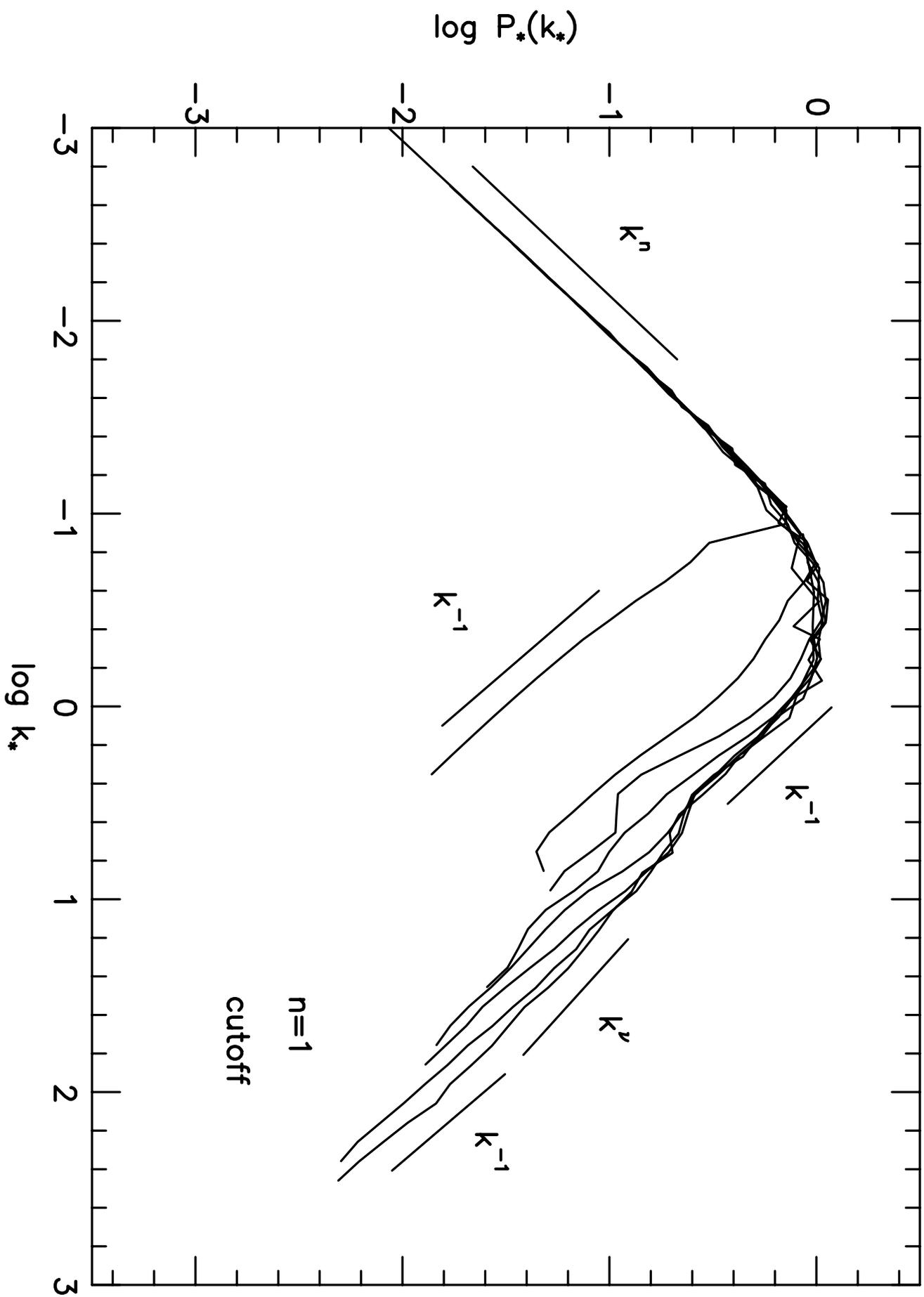
Fig.6.- The scaled power spectrum $P_*(k_*)$.
The initial power spectrum obeys the power-law with the index $n=1$ 
and it has a cutoff scale at $k_{cut}=127$. 

\noindent
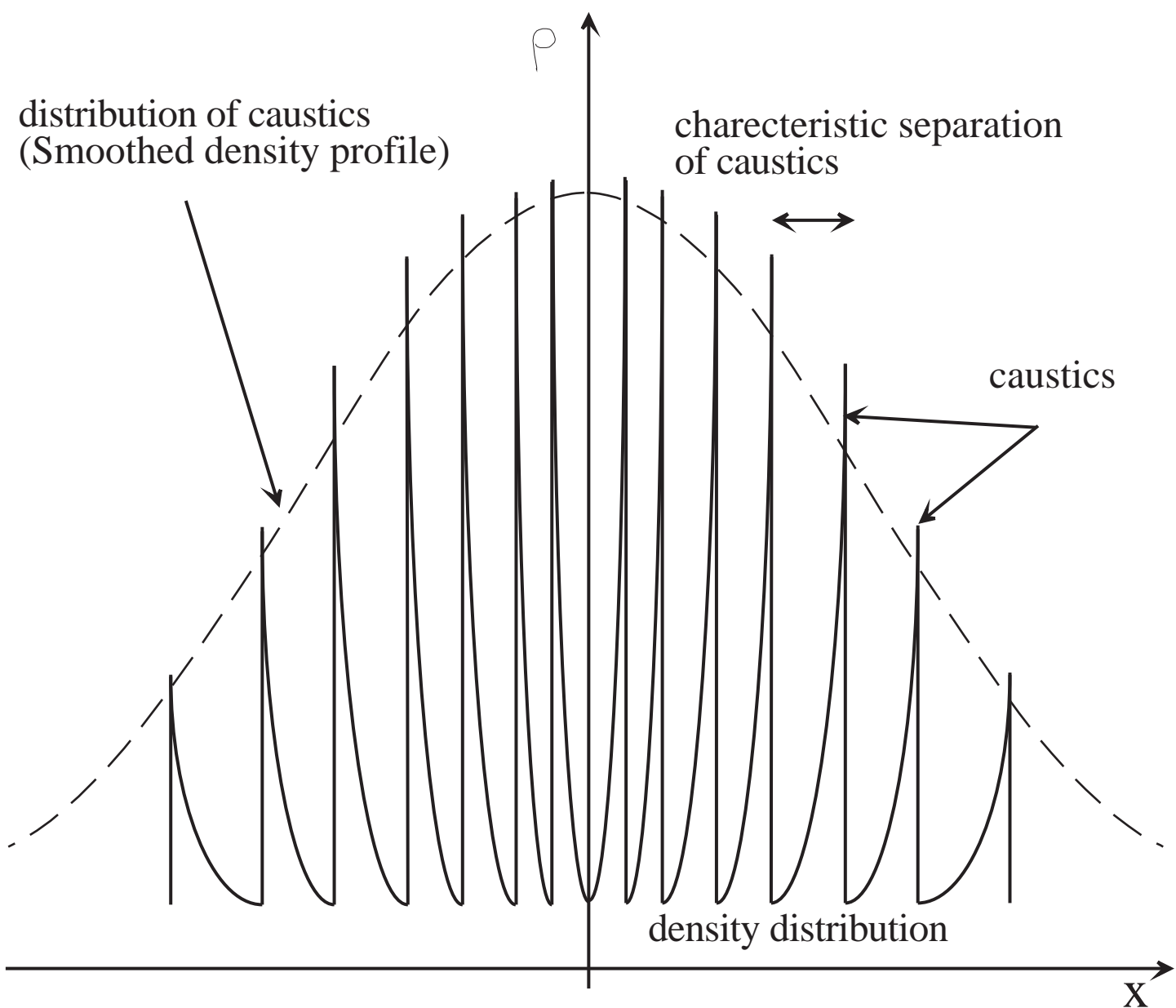
Fig.7.- The schematic of the density distribution in the single-wave case.
This figure shows that 
the density profile around one caustic and the distribution
of caustics, which determine the smoothed density profile on scales larger 
than the characteristic separation of caustics.

\noindent
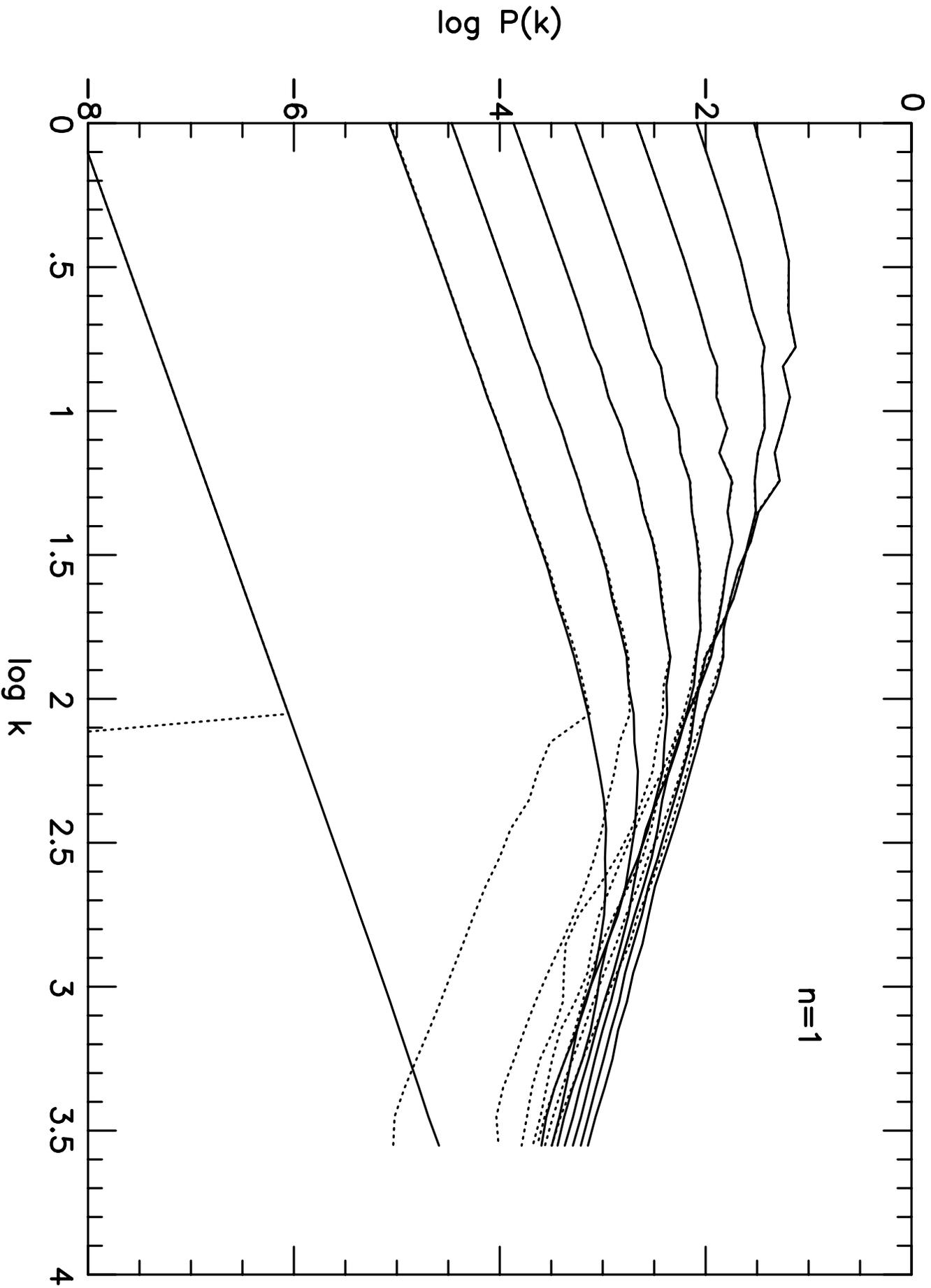
Fig.8.- Comparison of the evolution of the 
power spectrum between the scale-free case (solid line) 
and the cutoff-case (dashed line). 
Both the solid curves and dashed curves shown are (bottom to top) 
at $a=0.05,1,2,4,8,16,32,64$.
The initial power spectrum obeys the power law with the power index $n=1$
both for the scale-free case and for the cutoff-case.
For the cutoff-case, 
the initial power spectrum has a cutoff at $k_{cut}=127$.  

\noindent
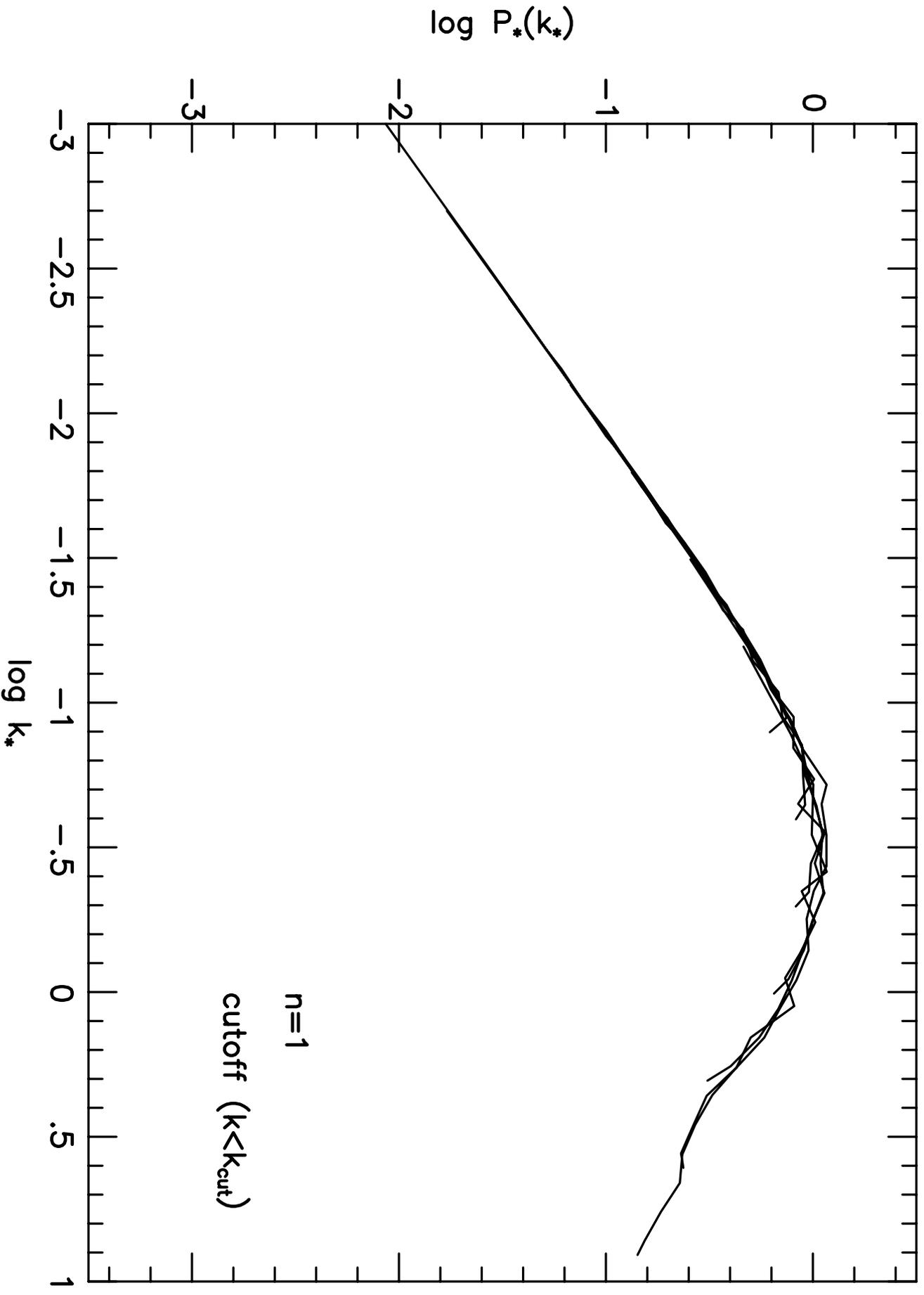
Fig.9.- The scaled power spectrum at a=1,2,4,8,16,32,64 on scales larger 
than the cutoff scale.  The initial 
power spectrum obeys the power law with the power index n=1 and
it has a cutoff scale at $k_{cut}=127$.

\noindent
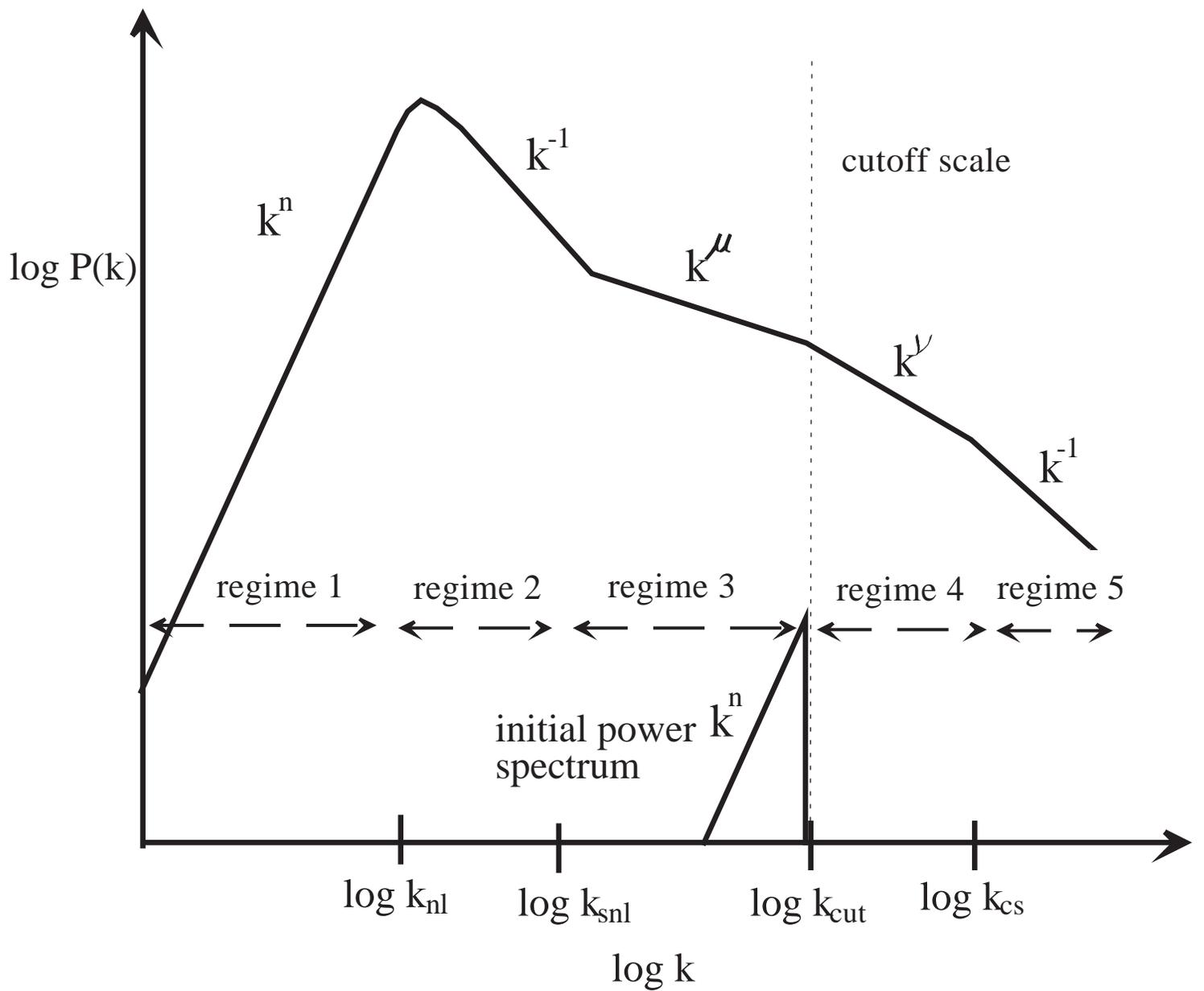
Fig.10.- The schematic of the power spectrum 
with a general initial power spectrum
at a certain time after the first appearance of caustics. 
\end{document}